\documentclass[global,twocolumn]{svjour}
\usepackage{graphicx}
\usepackage{array}
\usepackage{SIunits}
\usepackage{color}
\usepackage{dcolumn}

\def\ket0{$\left|0\right>$}
\def\ket1{$\left|1\right>$}
\def\ca40{$^{40}\mathrm{Ca}^+$}
\def\T2{$\mathrm{T_2$}}
\def\Pr3{$\mathrm{Pr^{3+}}$}
\def\ket#1{$\left|#1\right>$}

\def\eq#1#2{\begin{equation}\label{Eq:#1}#2\end{equation}}
\def\eref#1{(\ref{Eq:#1})}
\def\fref#1{\ref{Fig:#1}}

\begin{document}

\title{Precision measurements in ion traps using slowly moving standing waves}

\author{A.~Walther, U.~Poschinger, K.~Singer, F.~Schmidt-Kaler}
\institute{Institut f\"ur Quantenphysik, Universit\"at Mainz, Staudingerweg 7, 55128 Mainz, Germany}

\date{\today}
\maketitle

\begin{abstract}
The present paper describes the experimental implementation of a measuring technique employing a slowly moving, near resonant, optical standing wave in the context of trapped ions. It is used to measure several figures of merit that are important for quantum computation in ion traps and which are otherwise not easily obtainable. Our technique is shown to offer high precision, and also in many cases using a much simpler setup than what is normally used. We demonstrate here measurements of i) the distance between two crystalline ions, ii) the Lamb-Dicke parameter, iii) temperature of the ion crystal, and iv) the interferometric stability of a Raman setup. The exact distance between two ions, in units of standing wave periods, is very important for motional entangling gates, and our method offers a practical way of calibrating this distance in the typical lab situation.
%
\end{abstract} 

\section{Introduction}
\label{sec:Intro}
\begin{figure}[ht!]
		\begin{center}
    \includegraphics[width=6cm]{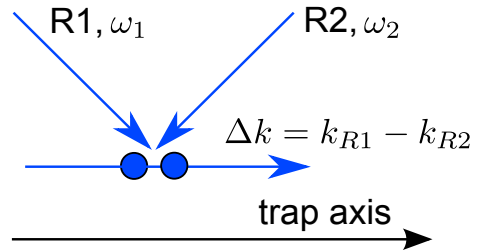}
    \caption{ Alignment of the two components of the beams together with the resulting direction of the $\mathrm{\Delta k}$-vector (given by $\mathrm{k_{R1} - k_{R2}}$) with respect to the trap axis.}
    \label{Fig:setup}
    \end{center}
\end{figure}

During the past decade ion traps have been shown to be one of the most promising candidates for the implementation of quantum computers~\cite{Blatt2008} and quantum simulators~\cite{JOHANNING2009b}. A basic building block for these two scientific endeavours is the controlled interaction of several ions with an optical standing wave. In the case of quantum computing, a tool for creating the necessary entangled states is the controlled motion of the ions that will couple via the Coulomb interaction, such as in the case of the geometric phase gate~\cite{LEIBFRIED2003a}. In this type of gate, a specific collective vibrational mode is chosen to mediate the coupling, high fidelity interactions is possible only with precise control of the experimental situation. This includes good knowledge of parameters such as the alignment of the ions in the standing wave and their localization (ion crystal temperature), in order to reside within the Lamb-Dicke regime where the ions are localized better than the wavelength of the standing wave. These parameters can be very challenging to actually measure. Fig. \ref{Fig:setup} depicts a typical experimental situation with an ion crystal in a standing wave from two laser beams irradiated from outside the apparatus. The actual wavelength of the beat pattern along the trap axis depends on the angles of the beams with respect to this axis and is therefore hard to set a priori. As the ion distance in a typical experiment is $10^1 - 10^2$ times the beat pattern period, an alignment precision of better than 1\% is clearly required.
Here we present a new technique, which is based on a slowly moving standing standing wave, designed to measure several important parameters for controlling entangling operations in ion traps. We show measurements of i) the distance between two crystalline ions, ii) the Lamb-Dicke parameter, iii) the temperature of the ion crystal, and iv) the interferometric stability of the Raman laser beam setup that is used to excited the ion motion. The basic idea of the standing wave technique is to use two laser beams, tuned close to a resonance of a dipole transition that can both cool the ions and yield fluorescence, but with a very small frequency offset with respect to each other, in our case chosen to be 2~Hz. This creates a very slowly moving wave, in which ions in a peak of the wave yield fluorescence but ions in a valley does not. For a two ion crystal in particular, a strong 2~Hz fluorescence beat pattern will be detected only if both ions are in peaks of the wave simultaneously, and thus the amplitude of the beat pattern becomes a direct measurement of the distance between the two ions, in units of the wave period. The technique is shown to offer very good accuracy, allowing measurements of the Lamb-Dicke parameter in the sub-percent range. Looking at this from the opposite point of view, one can also say that the ions are used to map the properties of the light with a sub-wavelength precision, similarly as to what has been discussed before~\cite{Eschner2003}. From this perspective, we are operating in a range between the single ion resolution of~\cite{Eschner2003} and the completely averaged regime that is obtained for large ion crystals~\cite{Herskind2009}. The method also operates without ground state cooling, which together with the fact that the beating pattern that can be observed in real time, enables the method to be used as a practical ion distance calibration tool. In the context of probing condensed matter objects by light or particle scattering, our method represents a measurement of the structure factor and Debye-Waller factor of the ion crystal, although with a fixed wavenumber of the probing light and a structure that can be varied by means of external control parameters. This could be used as an ideal tool for studying implementations of the Frenkel-Kontorova model with trapped ions as proposed in Ref.~\cite{GARCIA2007}.

The paper is organized as follows. In Sec.~\ref{sec:Methods}, the technique is explained in detail, with the experimental setup described in Sec.~\ref{sec:Exp}, and with the results of measuring the different parameters presented in Sec.~\ref{sec:Results}. In Sec.~\ref{sec:Relevance}, we elucidate in detail why and how this technique can be useful for the optimization of entangling gate operations, finally in Secs. \ref{sec:generalization} and \ref{sec:conclusion} we discuss the suitability of our method as a standard measurement tool and a possible generalizations to more ions.

\section{Methods: Standing wave technique}
\label{sec:Methods}
For a single ion, the fluorescence intensity from two laser beams with the same polarization and with a detuning of $\Delta\omega$ with respect to each other is given by
\eq{I_1}{
	I_1(t) = A + B \cos\left(\Delta k_{\textrm{eff}} \cdot x - \Delta\omega t + \Delta \phi\right),
}
where $\Delta k_{\textrm{eff}}$ is projection of the difference k-vector between the two beams onto the trap axis and $\Delta \phi$ is their optical phase difference. The term $A$ is the background fluorescence signal, which will be ignored in the following. The amplitude of the temporally varying signal $B$ depends on the laser power and the specific saturation of the optical transition as well as on the overall detuning of the two beams relative to the line width. The laser power is chosen such that the transition is not saturated. 

The distance of two crystalline ions in a common harmonic potential characterized by the trap frequency $\omega_0$ is given by
\begin{equation}
2 l_0=\sqrt[3]{\frac{e^2}{4\pi\epsilon_0}\frac{2}{m\omega_0^2}}
\label{eq:spacing}
\end{equation}
For two ions with a center of mass position $x_0$, we can describe the beating component of the fluorescence as sum of the fluorescence rates from each ion:
\begin{eqnarray}
	I_2(t) & = &  I_1^{(1)}(t)+I_1^{(2)}(t) \nonumber \\
		& = & B \cos(\Delta k_{\textrm{eff}} \cdot (x_0+l_0) - \Delta\omega t + \Delta \phi) + \nonumber \\
		& & B \cos(\Delta k_{\textrm{eff}} \cdot (x_0-l_0) - \Delta\omega t + \Delta \phi),
\label{Eq:I_2}
\end{eqnarray}
which is implicitly dependent on $x_0$ and $l_0$. Note that this signal is not due to interference of fluorescence from the two ions \cite{EICHMANN1993}, which other imaging schemes use~\cite{Thiel2009}, but rather due to the fact that the fluorescence from the two ions is correlated or anticorrelated depending in their alignment with respect to the standing wave and that the detection does not provide spatial resolution. In the case of a detection apparatus providing spatially resolved detection of ions \cite{BURRELL2010}, cross correlation signals can be used in the same manner. 

We need to take into account that the ions are delocalized due to thermal and quantum fluctuations of their position. The probability of finding the ion at a center-of-mass position $x$ is given by a Gaussian probability distribution\footnote{Note that $\sigma_x=\sigma_0/\sqrt{2}$}:
\eq{Gaussian}{
	p_0(x) = \frac{1}{\sqrt{\pi \sigma_0^2}} e^{-\frac{(x-x_0)^2}{\sigma_0^2}},
}
where 
\eq{sigma}{
	\sigma_0 = \sqrt{\overline{n}_0+1/2}\sigma_0^{(0)}.
}
The width of the thermal distribution, $\sigma_0$, depends on the mean phonon number
\eq{nbar}{
	\overline{n}_0 = \frac{k_B T}{\hbar \omega_0},
}
and on the width of the ground state wavefunction, $\sigma_0^{(0)} = \sqrt{\hbar/2m\omega_0}$. From this it is clear that the width of the thermal packet depends in two ways on the axial trap frequency $\omega_0$: firstly because a change to the trap frequency signifies a change to the confinement which also changes the width of the ground state packet, and secondly because a different trap frequency also gives a different mean phonon number for a fixed amount of thermal energy. A similar argumentation holds for the fluctuations of the ion distance $l_0+l$ associated with the stretch mode vibrating at $\omega_1=\sqrt{3}\omega_0$. We now integrate over all fluctuations, such that the final fluorescence intensity of the beating component as a function of time and ion separation is given by
\eq{beating}{
	\bar{I_2}(t) = \int \! \! \! \int I_2(t) \, p_0(x') \, p_1(l') \, dx' \, dl',
}
where $I_2(t)$ is the two-ion intensity from Eq.~\eref{I_2} using the fluctuating coordinates with their corresponding probability distributions from Eq.~\eref{Gaussian}. This integral yields
\begin{figure*}[t]
		\begin{center}
    \includegraphics[width=17cm]{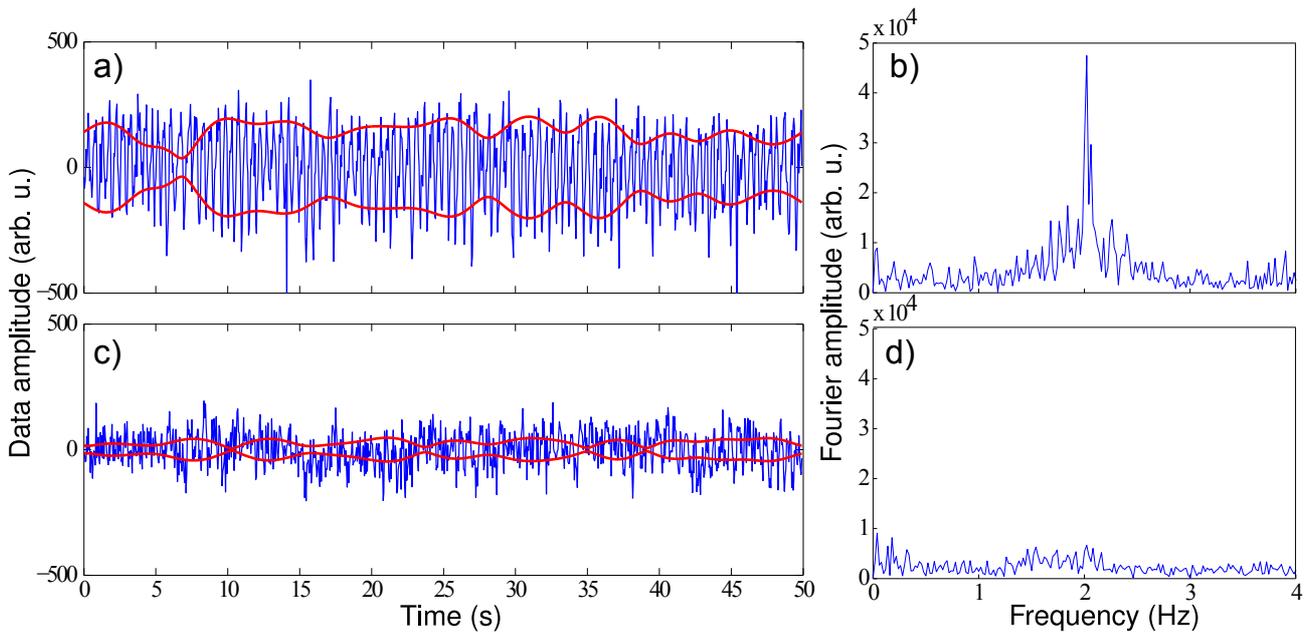}
    \caption{(color online) signal beating taken at two different ion distances, controlled by voltage on the segments of the microtrap. Part a) and c) show the raw data (blue curve) taken at a peak and a bottom of the oscillation strength respectively. Part b) and d) show the corresponding Fourier power spectra. The amplitude of the 2 Hz Fourier component has also been included in the left part of the figure in red. The strong oscillations correspond to the point at 4.82~\micro m in Fig.~\fref{distance} and the weak oscillations to the point at 4.73~\micro m.}
    \label{Fig:amplitude}
    \end{center}
\end{figure*}
\eq{I_t}{
	\bar{I}_2(t) \propto \cos(\Delta\omega t) \cos(\Delta k_{\textrm{eff}} l_0) e^{-(1/4)\Delta k^2_{\textrm{eff}}\left(\sigma_0^2 + \sigma_1^2\right)},
}
where $x_0=0$ and $\Delta\phi = \Delta\phi(t)$ represents phase instabilities between the interferometric branches. 
This expression shows that we obtain a signal on the beating frequency, $\Delta\omega$, which has an amplitude that depends on the ion separation. As can be seen, when the ion separation is exactly an integer number of wavelengths of the k-vector, the beat amplitude has a maximum, and for half-integer numbers, the beating disappears entirely. The beat amplitude also depends on the temperature of the ion crystal in the form of the wave packet width from the two different vibrational modes, $\sigma_0$ and $\sigma_1$.

\section{Experimental setup}
\label{sec:Exp}
The ions used in these experiments were \ca40, trapped in a segmented micro-structured Paul trap \cite{SCHULZ2006}, with radial trap frequencies of of $\omega_{\textrm{rad}}/2\pi = \{2.4, 3.0\}$~MHz. The axial dc trapping voltage is varied during the experiment in order to change the distance between the ions. This causes the axial trap frequency $\omega_0$ to vary between $2 \pi \cdot 0.96$ and $2 \pi \cdot 1.4$~MHz. The main laser interaction during the experiment consists of a pair of beams at 397~nm, both linearly polarized, hitting the ion at approximately 90$^{\circ}$ angle with respect to each other, as indicated by Fig.~\fref{setup}. 
These beams have a small detuning from each other of $\Delta\omega = 2\pi \cdot 2$~Hz, controlled by a waveform generator. They are also red-detuned slightly from the optical transition, such that they can provide the necessary Doppler cooling to the ions and serve as the source for the fluorescence that is detected. In addition, there are also lasers on 866 and 854 nm for the repumping purposes. The beating fluorescence light can be monitored in real time by a CCD camera, but for technical reasons, a photo multiplier tube was used to capture the data presented in Sec.~\ref{sec:Results}. A binning time of 50 ms was used, and the measurement durations for each run was 50~s. There is also an external magnetic field aligned at 45$^{\circ}$ with respect to the trap axis for the purpose of breaking the degeneracy of the ground level spin states.

\section{Results}
\label{sec:Results}
Measurements of the fluorescence were taken at a total of 33 different ion distances, with 5 runs of 50 seconds detection time for each distance. Examples of such a runs, when the ion distance was corresponding to a peak and a valley of the beating respectively, are displayed in Fig.~\fref{amplitude}. For proper analysis of the beating, we perform a Fourier transform of the signal. The Fourier spectrum is then multiplied with a supergaussian window
\eq{supergaussian}{
	g(\omega') = e^{-\frac{(\omega' - \Delta\omega)^{2 n}}{2 \sigma^{2 n}}},
}
where the center frequency is chosen to only keep the beating component, $\Delta\omega = 2\pi \cdot 2$~Hz, and where the order was chosen to be $n = 4$, giving a window that is roughly flat around the interesting parts, and zero elsewhere. The FWHM width of the window was chosen to be 0.3~Hz, but other widths was tested and the result was found to be relatively independent of the choice of window width. If one multiplies the window function to the positive side of the Fourier plane only, then upon inverse Fourier transform, a complex function, $A(t) e^{-i \phi(t)}$, is obtained. Here, $A(t)$ corresponds to the amplitude of the 2 Hz component of the signal as a function of time, and is plotted as the red overlayed curve in Fig.~\fref{amplitude}. Similarly, the phase factor, $\phi(t)$ corresponds to the phase evolution of the 2 Hz component and it plotted in Fig.~\fref{phase}.
\begin{figure}[ht]
    \includegraphics[width=8.5cm]{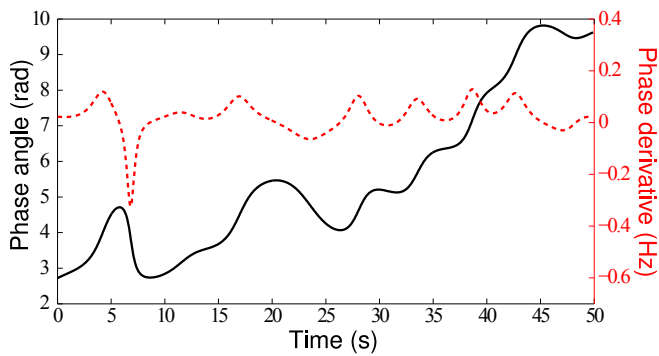}
    \caption{(color online) phase angle of the 2~Hz beating component as a function of time is plotted as the black (solid) line with corresponds axis to the left. Its derivative is plotted in red (dashed), with axis to the right, which gives a measurement of the frequency stability of the interferometric setup.}
    \label{Fig:phase}
\end{figure}
This figure also contains the derivative of the time dependent phase angle, which gives a measure of the frequency stability between the two arms of the interferometric setup. From an average of the standard deviation of the derivative for several traces we find an interferometric coherence time of about $\tau \sim 12$~s. The interferometric stability is important for other types of experiments, as this pair of laser beams is the same as is used to performed Raman type quantum gates in the quantum computer scheme, and we see that the coherence time is much longer than the typical gate times, which are in the order of 10 to 50 \micro s. We therefore expect that phase fluctuations during such gate operations are much less than $\pi/100$, thus not limiting the gate fidelity. Interferometer stability might also be important when applying optical light forces with fixed phase relations to other forces, such as electric drive forces from the trap electrodes, though this was not used in the context of this paper.
\begin{figure}[ht]
    \includegraphics[width=8.5cm]{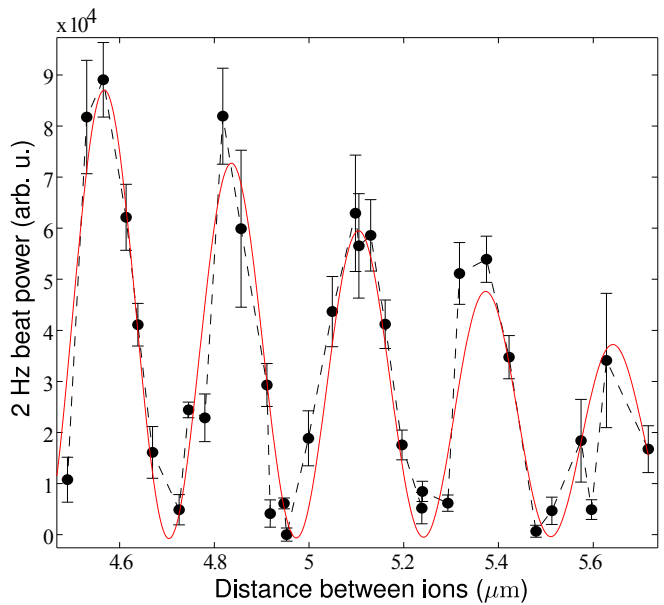}
    \caption{(color online) shows the measured beat power of the 2~Hz component of the fluorescence as a function of the ion distance, together with a theoretical fit using Eq.~\eref{I_k}. From the oscillation period, the effective wavelength of the beat pattern can be determined to $\lambda = 267.8(2)$~nm. Similarly, the decay of the oscillations for larger distances (lower trap frequencies) gives an ion crystal temperature of about 3.7)~mK.}
    \label{Fig:distance}
\end{figure}

Since the Fourier operation has picked out only the amplitude part of the beating, we can now write Eq.~\eref{I_t} as
\eq{I_k}{
	\bar{I}_2(t) \propto \cos(\Delta\omega t)\cos(\Delta k_{\textrm{eff}} l_0) e^{-(1/4)\Delta k_{\textrm{eff}}\left(\sigma_0^2 + \sigma_1^2\right)}.
}
In Fig.~\fref{distance}, the intensity of the beating component has been plotted as a function of ion distance for 33 difference distance. Eq.~\eref{I_k} has then been used to make a fitting to the result using the overall amplitude scaling together with $\Delta k$ and the ion crystal temperature, which gives both $\sigma_x$ and $\sigma_l$, as fitting parameters. Despite the sizable vertical error bars for each data point, the wave period can be determined with sub-percent accuracy, yielding a value of the standing wave wavelength of $\lambda = 2 \pi/\Delta k_{\textrm{eff}} = 267.8(2)$~nm. Note that the outstanding accuracy comes from the fact that there are many wave periods between the ions, and as such, even a small change to the wavenumber will be multiplied and change the fit substantially. The temperature was found to be $T = 3.7(2)$~mK, where the accuracy is in the 5\% range. Using this temperature, the common and stretch mode vibrational quantum numbers can be determined from Eq.~\eref{nbar}, and we find $\overline{n}_0 \approx 56(4)$ and $\overline{n}_1 \approx 32(2)$, for a trap frequency of $\omega_0 = 1.24$~MHz, which corresponds to the same ion distance that was chosen for Figs.~\fref{amplitude} and~\fref{phase}. These values are consistent with the expected results of Doppler cooling and with measurements using other techniques~\cite{Poschinger2010}. Furthermore, for this trap frequency and with the value of $k$ above, we calculate the Lamb-Dicke parameter to be $\eta = k_{\mathrm{eff}} \sqrt{\frac{\hbar}{2 m_{\mathrm{Ca}} \omega_0}} = 0.237$.

\section{Relevance for High Fidelity Entangling Gates}
\label{sec:Relevance}
The main motivation for the determination of the ion distance in units of the standing wave period is the requirement for high-fidelity quantum gates. The most commonly used entangling gates are based on the application of forces mediated by a spatially and temporally varying laser beat pattern, where the sign of the force can depend on the joint internal state of the ions. Two widely used gate schemes are the geometric phase gate \cite{LEIBFRIED2003a} and the Molmer-Sorensen gate \cite{SORENSEN2000,BENHELM2008b,ROOS2008b}. In the first scheme, the ion motion is near-resonantly driven by an oscillating ac-Stark shift arising from an off-resonant laser field; in the second scheme, simultaneous driving of red and blue sideband transitions gives rise to an effective entangling spin-spin interaction. In both cases, the driven collective vibrational mode is displaced and restored to the origin after the gate time, leading to the pickup of a geometric phase which is conditional 
on the joint internal state of the ions. The relevance of a precise alignment of the ions in the driving field is illustrated in the following for the case of the geometric phase gate driven by a bichromatic off-resonant laser field:
The classical force on one specific ion, $i$, is given by its position, $x$, in the standing wave:
\begin{equation}
F_{i}(x_i)=A\sin\left(\Delta k_{\textrm{eff}} x_i-\delta t+\Delta\phi\right)m_F^{(i)},
\end{equation}
where $m_F^{(i)}$ is the spin quantum number for ion $i$ ($m_J$ for species without nuclear spin), $\Delta k_{\textrm{eff}}$ is the effective wavevector along the trap axis, 
and $\Delta\phi$ is the phase difference of the two beams. $\Delta\omega$ is the difference between the two-photon detuning and the vibrational frequency, 
i.e. the effective detuning of the drive from the collective vibrational frequency. The total force driving the STR mode of a two ion crystal is the given by the difference of the individual forces acting on the ion, which is due to the fact that the forces on the ion need to be opposite for driving the STR mode. The spin configuration of the crystal in conjunction with the ion distance now determines the acting drive force, e.g. if the laser beams are tuned to drive the STR mode, \textit{only} the even spin configurations (\ket{\uparrow\uparrow} and \ket{\downarrow\downarrow}) are driven by the laser wave if the ion spacing (Eq. \ref{eq:spacing}) corresponds \textit{exactly} to a half integer multiple of drive period, $2l_0=(n+1/2)\lambda_{\textrm{eff}}$. By contrast, \textit{only} the odd components are displaced if the spacing is \textit{exactly} an integer multiple of the drive period. For a conditional phase gate, the drive amplitude such that the desired differential geometric phase between even and odd state is picked up, $\Delta\Phi=\Phi_{\textrm{even}}-\Phi_{\textrm{odd}}$. In the case of nonideal ion spacing, corresponding drive amplitudes and durations can still be found if the difference of the forces acting on odd and even states is large enough. However, nonideal ion spacing will lead to a larger required pulse area. The minimum possible amount of laser power can therefore only be used upon precise calibration of the ion spacing. Any increased amount of laser power will lead to increased decoherence by phase and intensity fluctuations. For off-resonant gate interactions, stimulated Raman scattering will lead to a decoherence proportional to the total employed pulse area \cite{OZERI2005,OZERI2007} even in the limit of technical perfection. The precise distance calibration becomes even more important for fast gates, when stronger drive amplitudes are used and the displacements become large enough such that effects beyond the Lamb-Dicke regime have to be considered \cite{MCDONNELL2004,Poschinger2010}. Here, the precise knowledge of the driving strengths is important to chose the gate time such that the displacement is completely undone at the return time and no information about the spin is left in the motional degree of freedom. Envisaged gate schemes for larger ion numbers enabling quantum computing in decoherence free subspaces have more stringent requirements on the crystal alignment \cite{IVANOV2010}, which will be covered in the next section. Furthermore, closed loop strategies such as optimal control theory might turn out to be helpful for the realization of quantum gates \cite{TIMONEY2008}, however, they require the system parameters as input such that the knowledge of the effective wavelength is would even be of crucial importance. 

\section{Generalization to an arbitrary ion number}
\label{sec:generalization}
We now consider the magnitude of the fluorescence oscillation signal for a linear ion crystal consisting of an arbitrary number, $N$, of ions. The position of the $i$-th ion is given by 
\begin{equation}
x_i=x_i^{(0)}+\sum_{j=1}^N a_{ij} q_j 
\end{equation}
with the equilibrium positions $x_i^{(0)}$, the normal coordinates $q_j$ pertaining to the $N$ axial normal modes and the Hessian matrix $a_{ij}$, see Ref. \cite{JAMES1998}. The oscillatory component of the fluorescence signal for ion $i$ is given by 
\begin{eqnarray}
I_n^{(i)}&=&\cos(\Delta\omega t)\cos\left(\Delta k_{\textrm{eff}} x_i\right) \nonumber \\
&=&\textrm{Re}\left(e^{i\Delta k_{\textrm{eff}} x_i^{(0)}}e^{i\Delta k_{\textrm{eff}}\sum_j a_{ij} q_j}\right),
\end{eqnarray}
where the reflection symmetry of the crystal about $x=0$ has been used and where $\Delta\phi=0$ as well as a homogeneous light field intensity has been assumed without loss of generality. Averaging over the thermal and quantum fluctuations yields
\begin{equation}
\bar{I}_n^{(i)}(\Delta\omega)=\cos(\Delta\omega t)\textrm{Re} \left(e^{i\Delta k_{\textrm{eff}} x_i^{(0)}}\prod_j f_{ij}\right),
\end{equation}
where 
\begin{eqnarray}
f_{ij}&=&\int_{-\infty}^{+\infty}dq_j e^{i\Delta k_{\textrm{eff}}\sum_j a_{ij} q_j-\frac{q_j^2}{2\sigma_j^2}} \nonumber \\
&=&e^{-(1/2)\Delta k_{\textrm{eff}}^2 a_{ij}^2 \sigma_j^2}.
\end{eqnarray}
\begin{figure}[ht]
    \includegraphics[width=9.5cm]{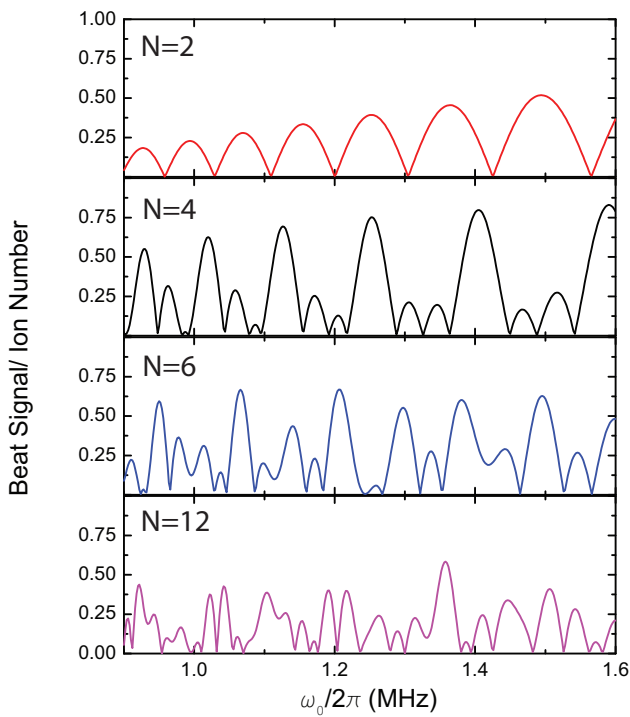} 
    \caption{(color online) Theoretical beat signal amplitudes versus trap frequency for linear ion crystals of different size. Note the periodic structure in the case of four ions, which can still be placed such that all spacings are integer multiples of the beat pattern period. This is not the case for larger crystals, as can be seen for the cases of six and twelve ions, where the irregularity increases towards larger ion numbers.}
    \label{Fig:theocurves}
\end{figure}
The total beat amplitude is the given by the sum over the ion's individual contributions $I_n(t)=\sum_i I_n^{(i)}(t)$. The total beat amplitude per ion versus 
trap frequency is plotted in Fig. \ref{Fig:theocurves} for different ion numbers. It shows that the individual ion localization is better for larger ion numbers. This counterintuitive behavior can be explained by the fact that for the low-frequency eigenmodes, the contributions the each ion's delocalization $a_{ij}$ are smaller. In case of the COM mode, this is simply due to the fact that the effective oscillator mass is $n$-times the mass of each ion. The high frequency modes are subject to a lower initial population according to Eq. \eref{nbar}, such that they do not contribute significantly to the delocalization at sufficiently low temperatures.  

\section{Conclusion and outlook}
\label{sec:conclusion}
We have presented a technique where the fluorescence from a two-ion crystal in a slowly moving standing wave was used to accurately measure the distance between ions in units of the standing wave wavelength. Being able to set the ion distance exactly to an integer number of the wavelength is important for maximizing the fidelity of quantum gate operations that are based on exciting motional interactions. In addition we show that this fluorescence signal also can be used to obtain the temperature of the ion crystal as well as information about the interferometric stability between the beams. Although a figure for the localization of ions in the standing wave, i.e. the ion temperature, is obtained, the method is not directly suited as a thermometry tool, as the temperature is determined in this case by the Doppler cooling exerted by the standing wave. The accuracy is limited to about 5\%, however the method can still be used to assert that the localization of the ions, i.e. the quality of the Doppler cooling, is slightly above what is expected from the Doppler cooling limit. Explanations for this might be competing Doppler heating effects on the micro-motion sidebands, or uneven cooling depending on sidebands and ion position (for further discussion on this, see~\cite{Cirac1992}). Since the technique is based on continuous laser interaction on a slow time scale, monitoring the signal can be done in real time without additional experimental obstacles like ground state cooling or driving coherent dynamics on motional sidebands. It therefore allows for direct alignment of ion crystals to the standing wave by fine adjustment of the trap frequency. In the future, we plan on applying this method for optimizing the properties of four-ion crystals prior to gate operations in decoherence-free subspaces~\cite{IVANOV2010}.


\bibliographystyle{unsrt}
\bibliography{Ref_lib_ions}

\end{document}